\documentclass{appolb}
\usepackage{epsfig}
\usepackage[switch]{lineno}
\usepackage{cite}

\begin{document}
\title{Probe the QCD phase diagram with $\phi$-meson
production in relativistic nuclear collisions at STAR%
\thanks{Presented at ``Strangeness in Quark Matter", Krak\'{o}w, Poland, September 18-24, 2011\\Email: xpzhangnju@gmail.com}
%
}
\author{Xiaoping Zhang (For the STAR Collaboration)
\address{Department of Engineering Physics, Tsinghua University,
Beijing 100084, China} } \maketitle
\begin{abstract}
We present $\phi$-meson transverse momentum distribution as well as
its elliptic flow ($v_{2}$) measurements in Au + Au collisions at
center-of-mass energy per nucleon pair $\sqrt{s_{NN}}$ = 7.7, 11.5
and 39 GeV with the data taken from STAR experiment at RHIC in the
year 2010. We discuss the energy dependence of $\phi$-meson elliptic
flow ($v_{2}$) and central-to-peripheral nuclear modification
factors ($R_{\textrm{\scriptsize{CP}}}$). The $v_2$ of $\phi$-mesons
are compared to those from other hadron species. The implications on
partonic-hadronic phase transition are discussed.
\end{abstract}

\section{Introduction}
The goals of the Beam Energy Scan program at BNL Relativistic Heavy
Ion Collider (RHIC) are to search for the Quantum Chromodynamics
(QCD) critical point as well as the phase transition boundary
between partonic and hadronic phases \cite{starbes}. By systematic
study of Au+Au collisions at $\sqrt{s_{NN}}$ = 5 $-$ 200 GeV, one
could access a wide region of temperature $T$ and baryon chemical
potential $\mu_{B}$ in the QCD phase diagram. Due to its small
hadronic cross section, the $\phi$-meson may decouple early from the
system, which makes it an important probe to study early partonic
evolution in high energy nuclear collisions \cite{nxuplot, phiprl,
na49phi}. In Au+Au collisions at RHIC with $\sqrt{s_{NN}}=200$ GeV,
the baryon-meson separation of $\phi$-mesons
$R_{\textrm{\scriptsize{CP}}}$ and Number-of-Constituent-Quark (NCQ)
scaling for $\phi$-meson $v_{2}$ at intermediate transverse momentum
($p_{T}$) have provided evidence for the formation of a new kind of
deconfined matter with partonic collectivity \cite{phiprl, philong}.
Because of the possible transition from partonic dominated phase to
hadronic dominated phase, it is expected that the established
paradigm for partonic degrees of freedom at top RHIC energy may
break at a given low collision energy. Especially, the absence or
reduction of collective flow and the breaking of NCQ scaling for
$\phi$-mesons could indicate the system in hadronic dominated phase
\cite{phincqb}. With the fully installed Barrel Time-of-Flight
detector (TOF) \cite{tof} in the year 2010, the signal-to-background
ratio of reconstructed $\phi$-mesons is greatly enhanced, which
improves the measurement of $\phi$-meson $v_2$ significantly at
lower energies. Here, we discuss the energy dependence of
$\phi$-meson $v_{2}$ and $R_{\textrm{\scriptsize{CP}}}$ in Au + Au
collisions at $\sqrt{s_{NN}}$ = 7.7, 11.5, and 39 GeV, based on data
taken from STAR experiment at RHIC in the year 2010.
\section{Experimental data analysis}
The Au+Au collision events collected by the minimum bias trigger are
used in this analysis. STAR's Time Projection Chamber (TPC)
\cite{tpc} and TOF \cite{tof} are used for tracking, particle
identification and event plane determination. The events are
required to have a primary Z vertex (along beam direction) within
40, 50, and 50 cm of the center of the TPC for Au+Au collisions at
$\sqrt{s_{NN}}$ = 7.7, 11.5, and 39 GeV, respectively, to ensure
nearly uniform detector acceptance. After the event selection, we
obtain about 4 million, 12 million, and 130 million Au+Au minimum
bias triggered events at $\sqrt{s_{NN}}$ = 7.7, 11.5, and 39 GeV,
respectively. For the $\phi$-meson $v_{2}$ analysis at 39 GeV, we
use the full statistics (130 million), while part of the statistics
(16 million) is used for spectra analysis at present. The collision
centrality is determined by the measured raw charged hadron
multiplicity from the TPC within a pseudorapidity window $|\eta|<$
0.5 \cite{glauber2}.

In $\phi$-meson spectra analysis, identified charged kaons, from the
TPC \cite{tpc}, are used to reconstruct the invariant mass
distributions of $\phi$-mesons at each well defined $p_{T}$ bin. For
$\phi$-meson $v_{2}$ analysis, additional kaon identification from
TOF \cite{tof} is used to get a clean kaon identification up to
$p_{T}=1.6$ GeV/$c$ and hence increase the signal-to-background
ratio of reconstructed $\phi$-mesons. The background is
reconstructed by a mixed event method. Raw yields of the $\phi$
mesons are determined by fitting the background subtracted invariant
mass distribution with a Breit-Wigner function superimposed on a
linear residual background function \cite{philong}. After taking
into account the detector resolution, the reconstructed $\phi$-meson
masses and widths are consistent with the PDG values \cite{pdg}.
Next, the corrections for reconstruction efficiency and geometrical
acceptance are applied to get the corrected $\phi$-meson yields. For
the $\phi$-meson $v_{2}$ analysis, we use the TPC $\eta$-sub event
plane method \cite{etasub}. With an $\eta$-gap between the charged
particles used in the event plane reconstruction and the measured
$\phi$-mesons, the non-flow effects related to the short range
$\eta$-correlation are expected to be reduced. This is especially
important when the expected $v_{2}$ value is small. More details of
STAR $\phi$-meson analysis can be found in References \cite{philong,
nasim}.

\section{Results and discussions}
\subsection{$\phi$-meson transverse momentum distribution and $R_{\textrm{\scriptsize{CP}}}$}
\begin{figure}[htbp]
\centering \vspace{-0.2cm}
\includegraphics[width=12.5cm]{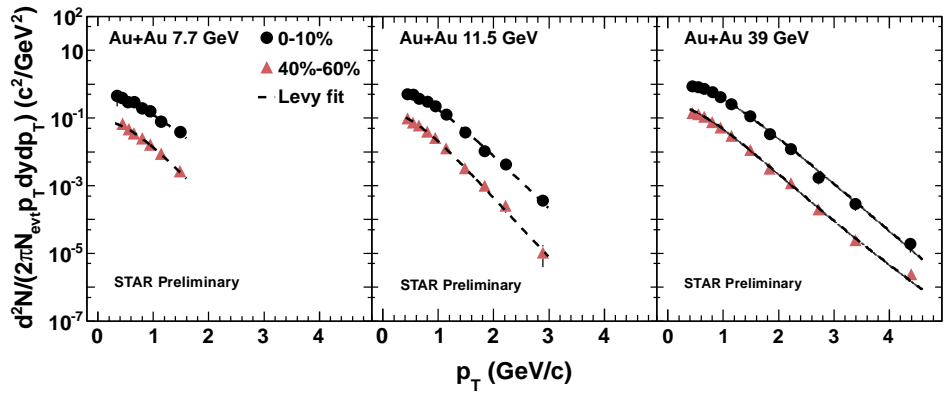}\vspace{-0.2cm}
 \caption{\label{fig1.fig} Transverse momentum spectra of mid-rapidity
($|y|<0.5$) $\phi$-mesons produced in Au+Au collisions at
$\sqrt{s_{NN}}$ = 7.7, 11.5 and 39 GeV (statistical error only). The
dashed curves are obtained from fits to the experimental data with a
Levy function \cite{philong}.}
\end{figure}

In Fig. \ref{fig1.fig}, we show the STAR preliminary $\phi$-meson
$p_{T}$ distribution in $\sqrt{s_{NN}}$ = 7.7, 11.5, and 39 GeV
Au+Au collisions at mid-rapidity ($|y|<0.5$). The 0-10\% in the plot
corresponds to central collisions, while 40\%-60\% to peripheral
collisions. The $p_{T}$ spectra in the above collision centralities
and energies are well described by a Levy function \cite{philong}.

\begin{figure}[htbp]
\centering \vspace{-0.2cm}
\includegraphics[width=6cm]{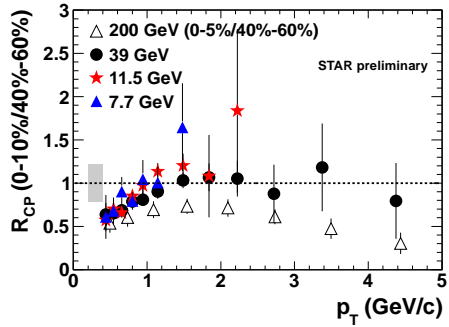}
\caption{\label{fig2.fig} $\phi$-meson
$R_{\textrm{\scriptsize{CP}}}$(0-10\%/40\%-60\%) at mid-rapidity
($|y|<0.5$) in Au+Au $\sqrt{s_{NN}}$ = 7.7, 11.5 and 39 GeV
collisions. $\phi$-meson
$R_{\textrm{\scriptsize{CP}}}$(0-5\%/40\%-60\%) in Au+Au 200 GeV
\cite{philong} is also plotted for comparison. Errors are
statistical only. The left grey band is the normalization error from
$N_{\textrm{\scriptsize{bin}}}$.}
\end{figure}

In Fig. \ref{fig2.fig}, we present the
$R_{\textrm{\scriptsize{CP}}}$ of $\phi$-mesons in Au+Au
$\sqrt{s_{NN}}$ = 7.7, 11.5, and 39 GeV collisions. The
$R_{\textrm{\scriptsize{CP}}}$ is defined as the ratio of particle
yields in central collisions over those in peripheral ones scaled by
the number of inelastic binary collisions
$N_{\textrm{\scriptsize{bin}}}$, that is,
\begin{eqnarray}
R_{\textrm{\scriptsize{CP}}}=\frac{[dN/(N_{\textrm{\scriptsize{bin}}}p_{T}dp_{T})]_{\textrm{central}}}{[dN/(N_{\textrm{\scriptsize{bin}}}p_{T}dp_{T})]_{\textrm{peripheral}}}.
\end{eqnarray}
Here, $N_{\textrm{\scriptsize{bin}}}$ is determined from Monte Carlo
Glauber model calculations \cite{glauber2}. The
$N_{\textrm{\scriptsize{bin}}}$ are $709.3\pm26.8$ ($76.0\pm17.0$),
$718.2\pm26.3$ ($76.3\pm16.3$) and $781.8\pm28.0$ ($80.3\pm17.7$)
for 0-10\% (40\%-60\%) Au+Au collisions at $\sqrt{s_{NN}}$ = 7.7,
11.5, and 39 GeV, respectively. The $R_{\textrm{\scriptsize{CP}}}$
will be unity if nucleus-nucleus collisions are just simple
superpositions of nucleon-nucleon collisions. Deviation of these
ratios from unity would imply contributions from nuclear or medium
effects.
At $\sqrt{s_{NN}}$ = 200 GeV Au+Au collisions, there is a large
suppression of $\phi$ meson yields in central collisions (0-5\%)
compared to peripheral collisions (40\%-60\%) \cite{philong}. This
could be interpreted as final-state partonic energy loss in dense
matter with a high gluon density \cite{partoneloss}. At
$\sqrt{s_{NN}}$ = 39 GeV Au+Au collisions, the
$R_{\textrm{\scriptsize{CP}}}$(0-10\%/40\%-60\%) is consistent with
unity for $p_{T} > 1$ GeV/$c$. This means no significant suppression
in Au+Au 39 GeV, which is quite different from Au+Au 200 GeV case.
We note that it does not mean that there is no nuclear or medium
effect, since there could be interplay between the Cronin effect
(parton $p_{T}$ broadening due to multiple scatterings) which
enhances hadron yields at intermediate $p_{T}$ \cite{Cronin1,xzhang}
and parton energy loss that suppresses the yields. However, the data
may suggest a smaller parton energy loss in Au+Au 39 GeV than that
in Au+Au 200 GeV. At 11.5 GeV, the $\phi$-meson
$R_{\textrm{\scriptsize{CP}}}$(0-10\%/40\%-60\%) seems to be larger
than 1 for $p_{T} > 1$ GeV/$c$.
In other words, the $\phi$-meson $R_{\textrm{\scriptsize{CP}}}$
reflects the decreasing partonic effects with decreasing beam
energies.
\subsection{$\phi$-meson elliptic flow $v_{2}$}
\begin{figure}[htbp]
\centering \vspace{-0.2cm}
\includegraphics[width=12.5cm]{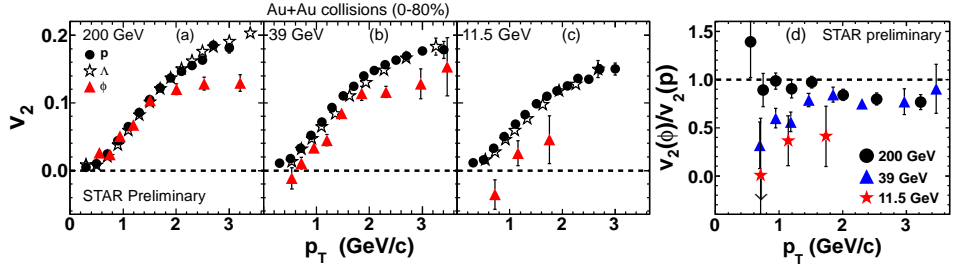}
\caption{\label{fig3.fig} (a)-(c): $\phi$-meson $v_{2}$ versus
$p_{T}$ compared to proton and $\Lambda$ in minimum bias Au+Au
collisions at 11.5, 39 and 200 GeV, respectively. The proton and
$\Lambda$ $v_{2}$ are from References \cite{v2200, plav2}. (d) The
ratio between $\phi$-meson $v_{2}$ and proton $v_{2}$ versus $p_{T}$
in above energies. The central value of $v_{2}(\phi)/v_{2}(p)$ is
far below 0 for the first low $p_{T}$ point in Au+Au 11.5 GeV
collisions. To suppress the y-axis scale, we place the central value
of this point at $v_{2}(\phi)/v_{2}(p)=0$ and use a down arrow to
show that the real central value of this point is below 0. Errors
are statistical only.}
\end{figure}

In Figs. \ref{fig3.fig}(a)-(c), we compare the $v_{2}$ of
$\phi$-meson to those of proton and $\Lambda$ in minimum bias Au+Au
collisions at $\sqrt{s_{NN}}$ = 11.5, 39, and 200 GeV. We use the
TPC event plane for $v_{2}$ measurement at 200 GeV \cite{v2200}, and
the TPC $\eta$-sub event plane at 11.5 and 39 GeV \cite{etasub} for
the above particles. One can see from Fig. \ref{fig3.fig}(a) that
$\phi$-meson $v_{2}$ is close to those of proton and $\Lambda$ for
$p_{T}$ below 1.5 GeV/$c$ in Au+Au 200 GeV collisions. This is
expected from the hydrodynamic model since the mass of $\phi$-meson
is close to those of the proton and $\Lambda$. For $p_{T}>2$
GeV/$c$, there is baryon-meson separation which could be understood
by quark coalescence in deconfined matter \cite{coal}.
In Au+Au 39 GeV collisions, we observe the similar baryon-meson
separation for $p_{T}>2$ GeV/$c$, indicating that there is a
partonic degree of freedom in the collision system. However, the
$\phi$-meson $v_{2}$ at low $p_{T}$ is smaller than the proton and
$\Lambda$ $v_{2}$ at 11.5 and 39 GeV. One can see from Fig.
\ref{fig3.fig}(d) that the ratios between $\phi$-meson $v_{2}$ and
proton $v_{2}$ show a decreasing trend with decreasing collision
energy at low $p_{T}$, indicating that relative to light quarks, the
strange quark collectivity becomes weaker. Possible reasons could be
that the strange quark are not fully thermalized in 11.5 and 39 GeV
Au+Au collisions and/or the hadronic interactions start to play an
important role in driving $\phi$-meson elliptic flow.

\section{Summary}
In summary, we present STAR preliminary measurements on $\phi$-meson
central-to-peripheral nuclear modification factor
($R_{\textrm{\scriptsize{CP}}}$) and elliptic flow ($v_{2}$) in
$\sqrt{s_{NN}}$ = 7.7, 11.5, and 39 GeV Au + Au collisions. The
$R_{\textrm{\scriptsize{CP}}}$ results show that there is no
significant suppression of intermediate-$p_{T}$ $\phi$-meson yields
in 0-10\% central Au+Au collisions at 39 GeV compared to peripheral
collisions (40\%-60\%), indicating a smaller partonic effect in
Au+Au 39 GeV collisions than in Au+Au 200 GeV collisions. The lack
of suppression in Au+Au 39 GeV collisions could be an interplay
between Cronin effect and parton energy loss. The baryon-meson
separation between $\phi$-meson and proton (or $\Lambda$) $v_{2}$ at
$p_{T}>2$ GeV/$c$ supports the quark coalescence picture in Au+Au 39
GeV collisions. At low $p_{T}$, the $\phi$-meson $v_{2}$ is smaller
than those of the proton and $\Lambda$, thus violates the mass
ordering observed in Au+Au 200 GeV collisions. This might be related
to the thermalization of $s$-quarks and/or the increasing
contributions from hadronic interactions to the elliptic flow with
decreasing collision energies.

\vspace{3ex}

X. Zhang thanks the support by the National Natural Science
Foundation of China (Grant Nos. 10865004, 10905029, 11035009,
11065005, and 11105079), by the China Postdoctoral Science
Foundation (Grant No. 20100480017), and by the Foundation for the
Authors of National Excellent Doctoral Dissertation of P.R. China
(FANEDD) (No. 201021).


\begin{thebibliography}{99}

\bibitem{starbes}  M.M. Aggarwal {\it et al.} (STAR
Collaboration.), arXiv:1007.2613.

\bibitem{nxuplot} H. van Hecke, H. Sorge, N. Xu, Phys. Rev.
Lett. \textbf{81}, 5764 (1998).

\bibitem{phiprl} B.I. Abelev {\it et al.} (STAR Collaboration.), Phys. Rev.
Lett. {\bf 99}, 112301 (2007).

\bibitem{na49phi} C. Alt {\it et al.} (NA49 Collaboration.), Phys. Rev. {\bf C78}, 044907 (2008).

\bibitem{philong} B.I. Abelev {\it et al.} (STAR Collaboration.), Phys. Rev.
{\bf C79}, 064903 (2009).


\bibitem{phincqb} B. Mohanty, N. Xu, J. Phys. \textbf{G36}, 064022 (2009); K.J. Wu, F. Liu, N. Xu, J. Phys. \textbf{G37}, 094029 (2010); J. Tian
{\it et al.}, Phys. Rev. \textbf{C79}, 067901 (2009).

\bibitem{tof} M. Shao et al., Nucl. Instrum. Meth. \textbf{A558}, 419
(2006); W.J. Llope, Nucl. Instrum. Meth. \textbf{B241}, 306 (2005).

\bibitem{tpc} K.H. Ackermann {\it et al.} (STAR
Collaboration), Nucl. Instrum. Methods \textbf{A499}, 624 (2003).

\bibitem{glauber2} B.I. Abelev {\it et al.} (STAR Collaboration), Phys. Rev. \textbf{C79}, 034909 (2009).

\bibitem{pdg} K. Nakamura {\it et al.} (Particle Data Group), J. Phys. \textbf{G37}, 075021
(2010).

\bibitem{etasub} B.I. Abelev {\it et al.} (STAR Collaboration), Phys. Rev. \textbf{C77}, 054901
(2008).

\bibitem{nasim} Md. Nasim {\it et al.} (STAR Collaboration), this conference.




\bibitem{partoneloss} M. Gyulassy, M. Plumer, Phys. Lett. \textbf{B243}, 432 (1990); X.N.
Wang, M. Gyulassy, Phys. Rev. Lett. \textbf{68}, 1480 (1992).

\bibitem{Cronin1} J.W. Cronin {\it et al.}, Phys. Rev. \textbf{D11}, 3105
(1975).

\bibitem{xzhang} X. Zhang {\it et al.}, Phys. Rev. \textbf{C84}, 031901(R)
(2011).
\bibitem{v2200} S. Shi {\it et al.} (STAR Collaboration), Nucl. Phys. \textbf{A830},
187c (2009).
\bibitem{plav2} A. Schmah {\it et al.} (STAR Collaboration), Quark
Matter 2011; S. Shi {\it et al.} (STAR Collaboration), this
conference.
\bibitem{hydro} P. Huovinen {\it et al.}, Phys. Lett. \textbf{B503}, 58 (2001).
\bibitem{coal} D. Molnar and S.A. Voloshin, Phys. Rev. Lett. \textbf{91}, 092301
(2003); R.C. Hwa, C.B. Yang, Phys. Rev. \textbf{C67}, 064902 (2003);
R.J. Fries {\it et al.}, Phys. Rev. Lett. \textbf{90}, 202303
(2003).
\end{thebibliography}
\end{document}